\documentclass[12pt,fleqn]{article}
\usepackage{latexsym,amsfonts,amssymb}
\newcommand{\be}[1]{\begin{equation}\label{#1}}
\newcommand{\ee}{\end{equation}}
\newcommand{\bs}[1]{\begin{equation}\label{#1}\arraycolsep=0em\begin{array}{l}}
\newcommand{\es}{\end{array}\end{equation}}
\newcommand{\bss}{\[\arraycolsep=0em\begin{array}{l}}
\newcommand{\ess}{\end{array}\]}
\newcommand{\ba}[1]{\begin{array}{#1}}
\newcommand{\ea}{\end{array}}
\newcommand{\bc}{\begin{center} }
\newcommand{\ec}{\end{center} }
\newcommand{\bt}[1]{\begin{tabular}{#1}}
\newcommand{\et}{\end{tabular} }

\renewcommand{\ge}{\geqslant}

\newcommand{\const}{\mathop{\rm const}\nolimits}

\newcommand{\dfrac}[2]{{\displaystyle\frac{#1}{#2}}}

\newcommand{\p}{\partial}
\newcommand{\R}{{\mathbb R}}
\newcommand{\CC}{{\mathbb C}}
\renewcommand{\Re}{\mathop{\rm Re}\nolimits}
\renewcommand{\Im}{\mathop{\rm Im}\nolimits}
\newcommand{\rank}{\mathop{\rm rank}\nolimits}

\begin{document}
\large

\begin{flushleft}
\bf Anatoly G. Nikitin~${}^\dag$ and  Roman O. Popovych~${}^\ddag$
\end{flushleft}

\begin{flushleft}
Institute of Mathematics, National Academy of Science of Ukraine,
\\ 3 Tereshchenkivs'ka Street, 01601, Kyiv-4, Ukraine\\
$\dag$~E-mail: nikitin@imath.kiev.ua\\
~~URL: http://www.imath.kiev.ua/\~{}nikitin/\\
$\ddag$~E-mail: rop@imath.kiev.ua\\
~~URL: http://www.imath.kiev.ua/\~{}rop/
\end{flushleft}

\vspace{2ex}

\noindent
{\Large\bf GROUP CLASSIFICATION\\[2mm] 
OF NONLINEAR SCHR\"ODINGER EQUATIONS}

\vspace{3ex}

\begin{abstract} 
The authors suggest a new powerful tool for solving group classification problems, 
that is applied to obtaining the complete group classification in the class of nonlinear 
Schr\"odinger equations of the form $i\psi_t+\Delta\psi+F(\psi,\psi^*)=0$.
\end{abstract}

\section{Introduction}

A nonlinear Schr\"odinger equation is one of the most interesting
and important models of contemporary mathematical physics.
Its completely integrable version was studied by many mathematicians (in particular, see
\cite{takhtadjian &faddeev.86.rus} and references therein).
This equation is also used in geometric optics~\cite{takhtadjian &faddeev.86.rus}
and nonlinear quantum mechanics~\cite{doebner&goldin.94}.

The main aim of the present paper is to perform the group classification
of nonlinear Schr\"odinger equations of the form
\be{nsche}
i\psi_t+\Delta\psi+F(\psi,\psi^*)=0
\ee
for a complex function $\psi=\psi(t,x)$ of $n+1$ 
real independent variables $t=x_0$ and $x=(x_1, x_2, \ldots, x_n)$, $n\ge1$,
with respect to an arbitrary element (a smooth function $F=F(\psi,\psi^*)$).
Here and below, a subscript of a function
means its differentiation with respect to the corresponding
 variable and the superscript .of any complex
quantity denotes complex conjugation. The indices $a$ and 
$b$ vary from 1 to $n$. It is assumed that summation
is carried out over repeated indices.

The class of equations (1) includes certain 
known equations as particular cases, namely, the free Schr\"odinger
equation ($F=0$), integrable Schr\"odinger equation with cubic nonlinearity ($F=\sigma|\psi|^2\psi$), 
Schr\"odinger equation with logarithmic nonlinearity ($F=\sigma\ln|\psi|\,\psi$,
 which, for  $\sigma\in\R$, is equivalent to the equation proposed in~\cite{bialynicki-birula&mycielski})), etc.

Until recently, the only method for the solution of problems of 
group classification of partial differential equations has been the direct
integration of defining equations for the coefficients of the operator of Lie symmetry
with subsequent cumbersome examination of all possible cases,
which considerably decreases the class of
solvable problems of group classification. Various 
modifications of the standard algorithm proposed, e.g., in ~\cite{zhdanov&lahno.jphys.a.1999,nikitin&wiltshire.conf99}
enable one to solve some of these problems. In the present paper, 
we use a new approach to the group classification
of equations~(\ref{nsche})  based on the investigation of the compatibility 
of the classifying system of equations with respect to an ``arbitrary element''.

\section{Kernel of main groups and equivalence group} 

Let the infinitesimal operator
\bss
Q=\xi^0\p_t+\xi^a\p_a+\eta\p_\psi+\eta^*\p_{\psi^*}
\ess
generate a one-parameter symmetry group of equation~(\ref{nsche}). 
(Here $\eta$~is a complex-valued function and $\xi^0$, $\xi^a$~are real-valued functions of
$t$, $x$, $\psi$, $\psi^*$.)
By using the infinitesimal criterion of invariance
\cite{ovsyannikov.rus,olver.rus} 
passing
to the manifold defined in the extended space by the system of 
equation~(\ref{nsche}) and its conjugate, and carrying out separation
with respect to unbound variables, we obtain the following 
defining equations for the coefficients of the operator $Q$:
\bs{det.eqs.for.nsche.1}
\xi^0_\psi=\xi^0_{\psi^*}=\xi^0_a=\xi^a_\psi=\xi^a_{\psi^*}=\eta_{\psi^*}=0, \quad \eta_{\psi\psi}=0,
\\[1ex]
\xi^a_b+\xi^b_a=0, \; a\not=b, \quad 2\eta_{a\psi}=i\xi^a_t, \quad 2\xi^a_a=\xi^0_t 
\es
(there is no summation over $a$ here) and
\be{det.eqs.for.nsche.2}
\eta F_\psi+\eta^* F_{\psi^*}+(\xi^0_t-\eta_\psi)F+i\eta_t+\eta_{aa}=0.
\ee
Integrating equations~(\ref{det.eqs.for.nsche.1}), 
we obtain the following expressions for the coefficients of the operator $Q$:
\bss
\xi^0=\xi^0(t), \quad \xi^a=\dfrac{1}{2}\xi^0_t(t)+\kappa_{ab}x_b+\chi^a(t), 
\\[1.5ex]
\eta=\eta^1(t,x)\psi+\eta^0(t,x), \quad 
\eta^1{:}=\dfrac{i}{8}\xi^0_{tt}(t)x_ax_a+\dfrac{i}{2}\chi^a_t(t)x_a+\zeta(t),
\ess
where $\kappa_{ab}=-\kappa_{ba}=\const,$ 
$\eta^0$ and $\zeta$ are complex-valued functions and $\chi^a$~a are real-valued functions.

Equation~(\ref{det.eqs.for.nsche.2})
 is a classifying condition that imposes further restrictions on the coefficients of the operator~$Q$.
depending on the form of the function $F$. If $F$
 is not fixed, then, upon the separation of equation~(\ref{det.eqs.for.nsche.2})
with respect to the ``variables'' $F$, $F_\psi$, $F_{\psi^*}$,
 we get $\eta=0$, $\xi^0_t=0$, $\chi^a_t=0$, 
which, with regard for relations~(\ref{det.eqs.for.nsche.1}),
yields the following statement:

\vspace{1ex}

\noindent
{\bf Proposition.} 
The kernel of main groups of equations of class~(\ref{nsche})
 is the Lie group whose Lie algebra $A^{\rm ker}$
 is the direct sum of Euclidean algebras in the space of the variable
$t$ and in the space of the variables $x$, i.e.,
\[
A^{\rm ker}=e(1)\oplus e(n)=\langle\p_t\rangle\oplus \langle\p_a, \; J_{ab}=x_a\p_b-x_b\p_a\rangle.
\]

\vspace{1ex}

The equivalence group of equation~(\ref{nsche})
coincides with the group generated by the collection of one-parameter
groups of local symmetries of the system
\be{system.equiv.nsche}
i\psi_t+\Delta\psi+F=0, \qquad F_t=0, \qquad F_a=0,
\ee
the infinitesimal operators of which have the form
\[
\widehat Q=\hat \xi^0\p_t+\hat \xi^a\p_a+\hat \eta\p_\psi+\hat \eta^*\p_{\psi^*}
+\hat \theta\p_F+\hat \theta^*\p_{F^*},
\]
where $\hat \theta$~is a complex-valued function of the variables $t$, $x$, $\psi$, $\psi^*$, $F$ 
and $F^*$, 
$\hat\eta$ is a complex-valued
function of the variables $t$, $x$, $\psi$, $\psi^*$, and 
 $\hat\xi^0$, $\hat\xi^a$~are real-valued functions of the variables 
$t$, $x$, $\psi$, $\psi^*$. 
Using the infinitesimal criterion of invariance for systems~(\ref{system.equiv.nsche})
and separating them with respect to un-bound
variables, we obtain defining equations for the coefficients of the operator $\widehat Q$. 
According to these equations, the Lie algebra of the equivalence group $G^{\rm equiv}$
of equation~(\ref{nsche}) is generated by the operators
\bs{basis.operators.of.nsche.equivalance.algebra}
\p_t, \quad \p_a, \quad J_{ab}, \\[1.5ex] 
t\p_t+\dfrac{1}{2}x_a\p_a-F\p_F-F^*\p_{F^*}, \quad \p_\psi+\p_{\psi^*}, \quad i(\p_\psi-\p_{\psi^*}), \\[1.5ex]
\psi\p_\psi+\psi^*\p_{\psi^*}+F\p_F+F^*\p_{F^*}, \quad i(\psi\p_\psi-\psi^*\p_{\psi^*}+F\p_F-F^*\p_{F^*}).
\es
Therefore, the equivalence transformations nontrivially acting on $F$ have the form
\be{general.equivalance.transformation.for.nsche}
\tilde t=\delta^2t, \quad \tilde x=\delta x, \quad 
\tilde \psi=\alpha\psi+\beta, \quad \tilde F=\delta^{-2}\alpha F,
\ee
where $\delta\in\R$, $\delta\not=0$, $\alpha,\beta\in\CC$, $\alpha\not=0$.

The restriction of the class of equations~(\ref{nsche}) can
lead to the appearance of equivalence transformations different
from transformations~(\ref{general.equivalance.transformation.for.nsche}) (see the proof).

\section{Result of classification} 

All the possible cases of extension of the maximal Lie invariance algebra 
of equation~(\ref{nsche}) are exhausted  
to within the equivalence transformations~(\ref{general.equivalance.transformation.for.nsche})
and their conditional extensions by the cases given in Tables 1 and 2. 
We present only basis elements from the complement with respect to $A^{\rm ker}$. 
After specifing the function $f$ in each case presented in Table~1, 
further extensions of the invariance algebra are possible, that is shown with Table~2. 
For describing results of classification, it is convenient 
to use the amplitude~$\rho=|\psi|$ and phase~$\varphi=\frac{i}{2}\ln\frac{\psi^*}{\psi}$
of the function~$\psi$. Let us introduce the notations
\bss
I:=\psi\p_\psi+\psi^*\p_{\psi^*}=\rho\p_\rho, \quad 
M:=i(\psi\p_\psi-\psi^*\p_{\psi^*})=\p_{\varphi}, \\[1.5ex] 
D:=t\p_t+\dfrac{1}{2}x_a\p_a, \quad G_a:=t\p_a+\dfrac{1}{2}x_aM, \\[1.5ex] 
\Pi:=t^2\p_t+tx_a\p_a-\dfrac{n}{2}tI+\dfrac{1}{4}x_ax_aM.
\ess

\noindent
{\normalsize {\bf Table 1.} 
Cases of extension where the expression for the function $F$
contains an arbitrary complex-valued smooth function 
$f$ of one real variable $\Omega$.
Here $\gamma$, $\gamma_1$, $\gamma_2$, $\delta$, $\delta_1$, $\delta_2$~are
real constants and $\theta=\theta(x)\!\in\!\R$~is an arbitrary solution of the 
equation $\:\Delta\theta=\delta_2\theta$.\par}

\bc\renewcommand{\arraystretch}{1.6}\tabcolsep=2mm 
\begin{tabular}{|p{7mm}|p{62mm}|p{21mm}|p{53mm}|}
\hline
&\hfil$F$\hfil&\hfil$\Omega$\hfil&\hfil Extension operators\hfil \\ \hline
$\hfil1.1$ & $f(\Omega)|\psi|^{\gamma_1}e^{\gamma_2\varphi}\psi$, $\:\gamma_1^2+\gamma_2^2\not=0$ & 
$|\psi|^{\gamma_2}e^{-\gamma_1\varphi}$ & $(\gamma_1^2+\gamma_2^2)D-\gamma_1I-\gamma_2M$
\\ \hline
$\hfil1.2$ & $(f(\Omega)+(\gamma-i)\delta\ln|\psi|)\psi$ & $|\psi|^\gamma e^{-\varphi}$ & $e^{\delta t}(I+\gamma M)$
\\ \hline
$\hfil1.3$ & $(f(\Omega)+\delta\varphi)\psi$, $\:\delta\not=0$& 
$|\psi|$ & $e^{\delta t}M$, $\;e^{\delta t}(\p_a+\frac{1}{2}\delta x_aM)$
\\ \hline
$\hfil1.4$& $f(\Omega)\psi$ & $|\psi|$ & $M$, $\;G_a$
\\ \hline
$\hfil1.5$& $f(\Omega)e^{i\psi}$ & $\Re\psi$ & $D+i(\p_\psi-\p_{\psi^*})$
\\ \hline 
$\hfil1.6$& $f(\Omega)+i(\delta_1+i\delta_2)\psi$ & $\Re\psi$ & 
$ie^{-\delta_1t}\theta(x)(\p_\psi-\p_{\psi^*})$ 
\\ \hline
\end{tabular}\ec

\vspace{1ex}

\newpage

\noindent
{\normalsize {\bf Table 2.} 
Cases of extension where the expression for the function $F$
does not contain arbitrary functions. Here 
$\gamma$, $\gamma_1$, $\gamma_2$, 
$\delta_1$, $\delta_2$, $\delta_3$, $\delta_4$~are
real constants, $\sigma\!\in\!\CC$, $\sigma\not=0$ (and $|\sigma|=1\bmod G^{\rm equiv}$);
$\eta^0=\eta^0(t,x)\!\in\!\CC$~is
an arbitrary solution of the original equation and $\theta=\theta(x)\!\in\!\R$~is an
arbitrary solution of the Laplace equation $\Delta\theta=0.$ For cases 2.9--2.15,
$\delta_j=\pm 1\bmod G^{\rm equiv}$
for one value of $j\!\in\!\{1;2;3;4\}$ if $\delta_j\not=0$.
\par}

\bc\renewcommand{\arraystretch}{1.6}\tabcolsep=2mm 
\normalsize\begin{tabular}{|p{8mm}|p{54mm}|p{83mm}|}
\hline
&\hfil$F$\hfil&\hfil Extension operators\hfil 
\\ \hline
$\hfill 2.1$ & 0 & $G_a$, $\:I$, $\:M$, $\:D$, $\:\Pi$, $\:\eta^0\p_\psi+{\eta^0}^*\p_{\psi^*}$
\\ \hline
$\hfill 2.2$ & $\gamma\psi+\psi^*$ & $I$, $\:\eta^0\p_\psi+{\eta^0}^*\p_{\psi^*}$
\\ \hline
$\hfill 2.3$ & $\sigma|\Re\psi|^\gamma,$ $\gamma\not=0,1$
 & $I+(1-\gamma)D$, $\; i\theta(x)(\p_\psi-\p_{\psi^*})$
\\ \hline
$\hfill 2.4$ & $\sigma\ln|\Re\psi|$& 
$I+D-i(t\Re\sigma+\frac{1}{2n}x_ax_a\Im\sigma)(\p_\psi-\p_{\psi^*})$, \hfill ${}$
$\;i\theta(x)(\p_\psi-\p_{\psi^*})$
\\ \hline
$\hfill 2.5$ & $\sigma e^{\Re\psi}$ & 
$D-\p_\psi-\p_{\psi^*}$, $\;i\theta(x)(\p_\psi-\p_{\psi^*})$
\\ \hline
$\hfill 2.6$ & $\sigma|\psi|^{\gamma_1}e^{\gamma_2\varphi}\psi$, \quad $\gamma_2\not=0$ & 
$M-\gamma_2D$, $\;\gamma_2I-\gamma_1M$
\\ \hline
$\hfill 2.7$ & $\sigma|\psi|^{\gamma}\psi$,
\quad $\gamma\not=0,\frac{4}{n}$& $G_a$, $\;M$, $\;I-\gamma D$ 
\\ \hline
$\hfill 2.8$ & $\sigma|\psi|^{4/n}\psi$, & $G_a$, $\;M$, $\;I-\frac{4}{n}D$, $\;\Pi$ 
\\ \hline
\multicolumn{3}{|c|}{In all cases below \quad $F=(-(\delta_1+i\delta_2)\ln|\psi|+(\delta_3-i\delta_4)\varphi)\psi$,\quad 
$\Delta=(\delta_2-\delta_3)^2-4\delta_1\delta_4$}
\\ \hline
$\hfill 2.9$ & $\delta_4=0$, $\delta_3\not=0$, $\delta_2\not=\delta_3$ & 
$e^{\delta_3t}M$, $\;e^{\delta_3t}(\p_a+\frac{1}{2}\delta_3x_aM)$, 
$\;e^{\delta_2t}(I-\frac{\delta_1}{\delta_2-\delta_3}M)$
\\ \hline
$\hfill 2.10$ & $\delta_4=0$, $\delta_3\not=0$, $\delta_2=\delta_3$ & 
$e^{\delta_3t}M$, $\;e^{\delta_3t}(\p_a+\frac{1}{2}\delta_3x_aM)$, 
$\;e^{\delta_2t}(I-\delta_1tM)$
\\ \hline
$\hfill 2.11$ & $\delta_4=0$, $\delta_3=0$, $\delta_2\not=0$ & 
$M$, $\;G_a$, $\;e^{\delta_2t}(\delta_2I-\delta_1M)$
\\ \hline
$\hfill 2.12$ & $\delta_4=0$, $\delta_3=0$, $\delta_2=0$, $\delta_1\not=0$ & 
$M$, $\;G_a$, $\;I-\delta_1tM$
\\ \hline
$\hfill 2.13$ & $\delta_4\not=0$, $\Delta>0$ & 
$e^{\lambda_it}(\delta_4I+(\lambda_i-\delta_2)M)$, $i=1,2$, \hfill ${}$
$\lambda_1=\frac{1}{2}(\delta_2+\delta_3-\sqrt{\Delta})$, 
$\lambda_2=\frac{1}{2}(\delta_2+\delta_3+\sqrt{\Delta})$
\\ \hline
$\hfill 2.14$ & $\delta_4\not=0$, $\Delta<0$ & 
$e^{\mu t}(\delta_4\cos\nu t\,I+((\mu-\delta_2)\cos\nu t-\nu\sin\nu t)M)$, \hfill ${}$
$e^{\mu t}(\delta_4\sin\nu t\,I+((\mu-\delta_2)\sin\nu t+\nu\cos\nu t)M)$, \hfill ${}$
$\mu=\frac{1}{2}(\delta_2+\delta_3)$, $\:\nu=\frac{1}{2}\sqrt{-\Delta}$
\\ \hline
$\hfill 2.15$ & $\delta_4\not=0$, $\Delta=0$ & 
$e^{\mu t}(\delta_4tI+\frac{1}{2}(\delta_3-\delta_2)tM+M)$, \hfill ${}$
$e^{\mu t}(\delta_4I+\frac{1}{2}(\delta_3-\delta_2)M)$, 
$\;\mu=\frac{1}{2}(\delta_2+\delta_3)$
\\ \hline
\end{tabular}\ec

\newpage

\section{Result of classification of the subclass {\mathversion{bold}$F=f(|\psi|)\psi$}} 

In the class of equation~(\ref{nsche}), 
we separate the subclass of Galilei-invariant equations with the nonlinearities $F=f(|\psi|)\psi$,
i.e., equations of the form
\be{ginsche}
i\psi_t+\Delta\psi+f(|\psi|)\psi=0.
\ee
The symmetry properties of these equations were 
studied in many papers (see, e.g., \cite{f.imath81,fs.jphA87.on.nsche,fshs.rus,chopyk.imath92}).
At the same time, we do not know works containing correct and exhaustive 
results concerning the group classification in the class
of equations~(\ref{ginsche}).
We separate them from the results presented in the previous section.

\vspace{1ex}

\noindent
{\bf Theorem.} 
The Lie algebra of the kernel of main groups of equations of class~(\ref{ginsche})
is the extended Galilei algebra
\[
A^{\rm ker}_{|\,|}=\tilde g(1,n)=\langle \p_t, \; \p_a, \; J_{ab}, \; G_a, \; M \rangle.
\]
The complete collection of inequivalent (with respect to local transformations) 
cases of extension of the maximal Lie invariance algebra of equations of the form~(\ref{ginsche})
is exhausted by the following cases 
(below, we present only basis operators from the complement to $A^{\rm ker}_{|\,|}$; 
$\sigma\in\CC$, $\sigma\not=0$, $|\sigma|=1\bmod G^{\rm equiv}$; 
$\gamma,\delta_1,\delta_2\in\R$; 
$\delta_2=\pm 1\bmod G^{\rm equiv}$ and $\delta_1=\pm 1\bmod G^{\rm equiv}$
 for cases (iii) and (iv), respectively]:

\vspace{0.5ex}

1. $f=\sigma|\psi|^\gamma,\:$ where $\:\gamma\not=0,\frac{4}{n}$: \quad $I-\gamma D$; 

\vspace{0.5ex}

2. $f=\sigma|\psi|^{4/n}$: \quad $I-\frac{4}{n}D$, $\;\Pi$;

\vspace{0.5ex}

3. $f=-(\delta_1+i\delta_2)\ln|\psi|,\:$ where $\:\delta_2\not=0$: \quad
$\!\!e^{\delta_2t}(\delta_2 I-\delta_1M)$;

\vspace{0.5ex}

4. $f=-\delta_1\ln|\psi|,\:$ where $\:\delta_1\not=0$: \quad $I-\delta_1tM$;

\vspace{0.5ex}

5. $f=0$: \quad $I$, $\;D$, $\;\Pi$, $\;\eta^0\p_\psi+\eta^0{}^*\p_{\psi^*},$ 

$\phantom{5.}$ where $\eta^0=\eta^0(t,x)$~is an arbitrary solution of the original
equation.

\section{Proof of the result of classification}
Let $A^{\rm max}=A^{\rm max}(F)$~be the maximal Lie invariance algebra of equation~(\ref{nsche})
with $F=F(\psi,\psi^*).$ 
If there is an extension (i.e., $A^{\rm max}\not=A^{\rm ker}$) 
then there exist such operators in $A^{\rm max}$
that the substitution of their coefficients in condition~(\ref{det.eqs.for.nsche.2}) 
gives a (nonidentical) equation for $F$. Each equation of this type has the form
\be{nshe.cc.for.F}
(a\psi+b)F_\psi+(a^*\psi^*+b^*)F_{\psi^*}+cF+d\psi+e=0,
\ee
where $a$, $b$, $c$, $d$, $e$ are complex constants. 
The differential consequences of equations of the form~(\ref{nshe.cc.for.F}),
which are of the first order (as differential equations), also reduce to the form~(\ref{nshe.cc.for.F}).
Thus, $A^{\rm max}\not=A^{\rm ker}$ iff the function $F$ satisfies $k$ ($k\in\{1;2;3\}$) 
independent equations of the form~(\ref{nshe.cc.for.F}). 
It is convenient to consider the linear case separately.

It should be noted that application of the standard methods of group classification in this problem reduces
to the investigation of different cases of integration of one equation of the form~(\ref{nshe.cc.for.F}) 
(depending on the values of the constants $a$, $b$, $c$, $d$, $e$) 
with subsequent decomposition into cases of further extension of the
symmetry group if the function $F$ satisfies certain additional conditions. 
This requires a cumbersome examination
procedure with multiple repetition of the same cases. 
The method proposed enables one to significantly decrease
the number of cases that should be examined.

\vspace{1ex}

\noindent
{\bf Linear case.} 
Let $F$ be a function linear in $(\psi, \psi^*)$, i.e., $F=\sigma_1\psi+\sigma_2\psi^*+\sigma_0$, 
where $\sigma_0$, $\sigma_1$, $\sigma_2$ are complex constants.
We can always set the constant $\sigma_0$ equal to zero by using the transformation 
$\tilde t=t$, $\tilde x=x$, $\tilde\psi=\psi+\nu_0+\nu_1t+\nu_2x_ax_a$
 from the extension $G^{\rm equiv}$; 
here, the complex constants $\nu_0$, $\nu_1$ and $\nu_2$ are
determined by the form of $F$. 
Then, depending on the value of the constant $\sigma_2$ 
($\sigma_2=0$ or  $\sigma_2\not=0$), by using
transformations from $G^{\rm equiv}$ and from the conditional extension of $G^{\rm equiv}$
($\tilde t=t$, $\tilde x=x$, $\tilde\psi=\psi e^{i\sigma_1t}$ or
 $\tilde\psi=\psi e^{-t\Im  \sigma_1}$)
the function $F$ can be reduced to cases 2.1 and 2.2 (where $\gamma=\Re\sigma_1$), 
respectively.

\vspace{1ex}

Below we assume that 
$F$ is a function nonlinear with respect to  $(\psi, \psi^*)$
and, hence, $(a,b)\not=(0,0)$ in (\ref{nshe.cc.for.F}).

\vspace{1ex}

\noindent
{\mathversion{bold}$k=1.$} It follows from~(\ref{det.eqs.for.nsche.2}) that
\[
\eta^1=\lambda a,\; \eta^0=\lambda b,\; \xi^0_t-\eta^1=\lambda c,\; 
i\eta^1_t+\Delta\eta^1=\lambda d,\; i\eta^0_t+\Delta\eta^0=\lambda e, 
\]
where $\lambda=\lambda(t,x)\in\R$, $\lambda\not=0$ (otherwise, $A^{\rm max}=A^{\rm ker}$). 

For $a\not=0$ $b=0\bmod G^{\rm equiv}$, whence $\eta^0=0$, $e=0$, $d=-i\delta a$, where $\delta\in\R$. 
In addition, if $c+a\not=0$ then $c+a=-|a|^2\bmod G^{\rm equiv}$, $\xi^0_{tt}=0$, $\chi^a_t=0$, 
$\zeta=\const$ and $d=0$ (case 1.1, where $\gamma_1=\Re a$, $\gamma_2=\Im a$). 
If $c+a=0$, $\Re a\not=0$, then $\Re a=1\bmod G^{\rm equiv}$, $\xi^0_t=0$, $\chi^a_t=0$, 
$\lambda_a=0$, $\lambda_t=\delta\lambda$ (case~1.2, where $\gamma=\Im a$).
If $c+a=0$, $\Re a=0$, then $\Im a\not=0$ (and, furthermore, $\Im a=1\bmod G^{\rm equiv}$), $\xi^0_t=0$, 
$\chi^a_{tt}=\delta\chi^a_t$, $\lambda=\frac{1}{2}\chi^a_tx_a+\Im\zeta$, $\Re\zeta=0$
(cases~1.3 and~1.4 for $\delta\not=0$ and $\delta=0$ respectively).

If $a=0$, then $b\not=0$ (and, furthermore, $b=i\bmod G^{\rm equiv}$), $\eta^1=0$ 
(whence, $d=0$, $\xi^0_{tt}=0$, $\chi^a_t=0$), $c\in\R$. 
Then, for $c\not=0$ $c=1\bmod G^{\rm equiv}$, $\lambda=\xi^0_t=\const$, 
$\eta^0=i\xi^0_t$, $e=0$, (case 1.5), 
and, for $c=0$ $\xi^0_t=0$, $\eta^0=ie^{-\delta_1t}\theta(x)$, where
$\Delta\theta=\delta_2\theta$, $\delta_1=\Re e$, $\delta_2=\Im e$ (case 1.6).

\vspace{1ex}

\noindent
{\mathversion{bold}$k=2.$} 
Assume that there exists a function $F$ nonlinear with respect to $(\psi, \psi^*)$
that satisfies a system of
two independent equations of the form~(\ref{nshe.cc.for.F}), i.e.,
\be{nshe.cc.for.F.2}
(a_j\psi+b_j)F_\psi+(a_j^*\psi^*+b_j^*)F_{\psi^*}+c_jF+d_j\psi+e_j=0,\quad j=1,2,
\ee
where $a_j$, $b_j$, $c_j$, $d_j$, $e_j$ ($j=1,2$)~are complex constants and
\[\rank\left(\ba{cccc}a_1&b_1&a_1^*&b_1^*\\a_2&b_2&a_2^*&b_2^*\ea\right)=2.\] 

\vspace{1ex}

\noindent
{\bf Lemma 1.} 
One of the following conditions is satisfied to within transformations from $G^{\rm equiv}$ and real
linear transformations of equations themselves:

\vspace{1ex}

1. $a_1=1$, $a_2=0$, $b_1=0$, $b_2=i$, $c_2=0$, $id_1=e_2(c_1+1)$, $d_2(c_1+2)=0$, 
$(c_1,e_1)\not=(0,0)$;
\vspace{1ex}

2. $a_1=1$, $a_2=i$, $b_1=b_2=0$, $d_1(c_2+a_2)=d_2(c_1+a_1)$, $c_1e_2=c_2e_1$;
 
\vspace{1ex}

3. $a_1=a_2=0$, $b_1=1$, $b_2=i$, $d_1c_2=d_2c_1$, $b_1d_2+c_1e_2=b_2d_1+c_2e_1$.
 
\vspace{1ex}

Equation~(\ref{det.eqs.for.nsche.2}), regarded as a condition on
$F$, is to depend on equations~(\ref{nshe.cc.for.F.2}) for any fixed operator from $A^{\rm max}$.
Therefore, for finding all defining equations for  $\xi^0$ and
 $\eta$ additional to~(\ref{det.eqs.for.nsche.1}), it suffices to equate to zero the
third-order minors of the extended matrix of the system of linear algebraic equations~(\ref{det.eqs.for.nsche.2}),
 (\ref{nshe.cc.for.F.2}) with respect to the ``unknowns'' $F_\psi$, $F_{\psi^*}$, $F$, i.e.,
\be{det.eqs.for.nsche.2a}\!\!\ba{l}
\left|\ba{ccc}
a_1\psi+b_1&a_1^*\psi^*+b_1^*&c_1\\a_2\psi+b_2&a_2^*\psi^*+b_2^*&c_2\\
\eta^1\psi+\eta^0&\eta^1{}^*\psi^*+\eta^0{}^*&\xi^0_t-\eta^1
\ea\right|=0,\\[5ex]
\left|\ba{ccc} 
a_1\psi+b_1&a_1^*\psi^*+b_1^*&d_1\psi+e_1\\a_2\psi+b_2&a_2^*\psi^*+b_2^*&d_2\psi+e_2\\
\eta^1\psi+\eta^0&\eta^1{}^*\psi^*+\eta^0{}^*&(i\eta^1_t+\Delta\eta^1)\psi+i\eta^0_t+\Delta\eta^0
\ea\right|=0.
\ea\ee
(Equations~(\ref{det.eqs.for.nsche.2a}) can be splitted with respect to the variables $\psi$ and $\psi^*$.)

We consider each case of Lemma 1 separately and seek only additional extensions (as compared with those
presented in Table 1) of the invariance algebra.

1. It follows from~(\ref{det.eqs.for.nsche.2a}) that $\eta^1\in\R$ (i.e., $\xi^0_{tt}=0$, 
$\chi^a_t=0$, $\zeta\in\R$), $\eta^0=i\rho(t,x)$, where $\rho\in\R$, 
$-\rho_t+i\Delta\rho+e_1\zeta+e_2\rho=0$, $i\zeta_t+d_1\zeta+d_2\rho=0$.
The additional extension $A^{\rm max}$  exists only if $d_1=d_2=e_2=0$, 
$c_1\in\R$, $c_1+1\not=0$. Under these conditions, equation~(\ref{nsche}) is reduced to case 2.3 
(if $c_1\not=0$), where $\gamma=-c_1$, or case 2.4 (if $c_1=0$), where $\sigma=-e_1$,
by the following transformation from the extension of $G^{\rm equiv}$:
$\tilde t=t$, $\tilde x=x$, $\tilde\psi=\psi+\nu_0+\nu_1t+\nu_2x_ax_a$, 
where the real constants $\nu_0$, $\nu_1$ and $\nu_2$ are determined by the form of $F$.

2. It follows from~(\ref{det.eqs.for.nsche.2a}) that $\eta^0=0$, 
$\tilde c_1\eta^1+\tilde c_2\eta^1{}^*=\xi^0_t-\eta^1$, 
$\tilde d_1\eta^1+\tilde d_2\eta^1{}^*=i\eta^1_t+\Delta\eta^1$, 
$\tilde e_1\eta^1+\tilde e_2\eta^1{}^*=0$, where 
\bs{definition.of.tilde.c1.c2.d1.d2.e1.e2}
\tilde c_1=\frac{1}{2}(c_1-ic_2), \quad 
\tilde d_1=\frac{1}{2}(d_1-id_2), \quad
\tilde e_1=\frac{1}{2}(e_1-ie_2), \\[1ex]
\tilde c_2=\frac{1}{2}(c_1+ic_2), \quad 
\tilde d_2=\frac{1}{2}(d_1+id_2), \quad
\tilde e_2=\frac{1}{2}(e_1+ie_2), 
\es
whence
$\tilde d_1(\tilde c_2+1)=\tilde d_2\tilde c_1$, 
$\tilde c_1\tilde e_2=\tilde c_2\tilde e_1$. 
System~(\ref{nshe.cc.for.F.2}) can be represented in the form
\[
\psi F_\psi+\tilde c_1F+\tilde d_1\psi+\tilde e_1=0,\quad 
\psi^* F_{\psi^*}+\tilde c_2F+\tilde d_2\psi+\tilde e_2=0.
\] 
$(\tilde c_1, \tilde c_2)\not=(0,0)$ (otherwise, we have a partial case of case 1.1).

If $\tilde c_1=-1$, $\tilde c_2=0$, then $\xi^0_t=0$, $\tilde e_1=\tilde e_2=0$ 
(otherwise, $A^{\rm max}=A^{\rm ker}$). Therefore,
\[
\chi^a_{tt}=(\delta_3-i\delta_4)\chi^a_t, \quad 
\zeta^1_t=\delta_2\zeta^1+\delta_4\zeta^2, \quad
\zeta^2_t=-\delta_1\zeta^1+\delta_3\zeta^2,
\]
where $\delta_1=\Re d_1$, $\delta_2=\Im d_1$, $\delta_3=-\Re d_2$, $\delta_4=\Im d_2$. 
Depending on the values of the constants~$\delta_l$, $l=\overline{1,4}$, we arrive at cases~2.9--2.15.

If $\tilde c_1=-1$, $\tilde c_2=0$, then the additional extension $A^{\rm max}$ exists only if 
$\tilde e_1=\tilde e_2=0$, $\tilde c_1+1=\tilde c_2^*\not=0$. Then, depending on the value of
$\tilde c_2$, by a transformation from the extension  $G^{\rm equiv}$,
one can reduce equation~(\ref{nsche}) (1) to case 2.6 (if $\tilde c_2\not\in\R$), where
$\gamma_1=-2\Re\tilde c_2$, $\gamma_2=-2\Im\tilde c_2$, 
or case~2.7 (if $\tilde c_2\in\R$, $\tilde c_2\not=-2/n$), whhere $\gamma=-2\tilde c_2$, 
or case~2.8 (if $\tilde c_2=-2/n$).

3. It follows from~(\ref{det.eqs.for.nsche.2a}) that $\eta^1=0$ (i.e., $\xi^0_{tt}=0$, 
$\chi^a_t=0$, $\zeta=0$), 
$\tilde c_1\eta^0+\tilde c_2\eta^0{}^*=\xi^0_t$, 
$\tilde d_1\eta^0+\tilde d_2\eta^0{}^*=0$, 
$\tilde e_1\eta^0+\tilde e_2\eta^0{}^*=i\eta^0_t+\Delta\eta^0$, where 
the constants $\tilde c_j$, $\tilde d_j$, $\tilde e_j$ ($j=1,2$) are defined in~(\ref{definition.of.tilde.c1.c2.d1.d2.e1.e2}).
We can represent system~(\ref{nshe.cc.for.F.2}) in the form
\[
 F_\psi+\tilde c_1F+\tilde d_1\psi+\tilde e_1=0,\quad 
F_{\psi^*}+\tilde c_2F+\tilde d_2\psi+\tilde e_2=0.
\] 

For the existence of an additional extension of $A^{\rm max}$, the following conditions are to be satisfied: 
$\tilde d_1=\tilde d_2=0$, $\tilde c_1^*=\tilde c_2\not=0$.
Hence, $\tilde c_1^*=\tilde c_2=-1\bmod G^{\rm equiv}$.
Then, by the transformation $\tilde t=t$, $\tilde x=x$, $\tilde\psi=\psi+it\Re e_1-\frac{i}{2n}x_ax_a\Im e_1$ 
from the extension of $G^{\rm equiv}$ equation~(\ref{nsche}) is reduced to case 2.5.

\vspace{1ex}

\noindent
{\mathversion{bold}$k=3.$}  The following statement is true:

\vspace{1ex}

\noindent
{\bf Lemma 2.} 
Suppose that a function $F$ satisfies a system of three independent equations of the form~(\ref{nshe.cc.for.F}).
Then the function $F$ is linear with respect to  $(\psi, \psi^*)$.

\vspace{1ex}

We have completed the classification of the class of equations~(\ref{nsche}).
In addition to the known particular cases presented
in~ \cite{fs.jphA87.on.nsche,chopyk.imath92}, 
we have obtained a complete collection of inequivalent equations~(\ref{nsche}) that admit a nontrivial
symmetry.

\vspace{1ex}

Note that the results concerning the group classification of systems of two equations of diffusion
(equation (\ref{nsche}) belongs to this class if it is regarded as a system of two equations for two real functions] 
were addused in~\cite{nikitin&wiltshire.conf99}. 
Our results confirm and improve the results presented in~\cite{nikitin&wiltshire.conf99}.

\medskip

{\bf Acknowledgments.} The authors are grateful to Dr.\ V.\ Boyko for useful discussions.

\end{document}